\documentclass[aps,pra,showkeys,twocolumn,superscriptaddress]{revtex4-2}
\usepackage[utf8]{inputenc}
\usepackage{amsmath,graphicx,array,amsthm}
\usepackage[colorlinks={true}]{hyperref}
\hypersetup{colorlinks=true,linkcolor=red,citecolor=magenta,urlcolor=blue}
\usepackage{indentfirst}
\usepackage{float}
\usepackage{braket}
\usepackage{diagbox}
\usepackage{amssymb}
\usepackage{bbold}
\usepackage{colortbl}
\usepackage{multirow}
\newcommand{\orcidlink}[1]{}

\newcommand{\CT}{C_{\mathrm{T}}}
\newcommand{\CC}{C_{\mathrm{C}}}
\newcommand{\CL}{C_{\mathrm{L}}}
\newcommand{\rhoI}{\rho_{\mathrm{I}}}
\newcommand{\piro}{\pi_{\rho}}

\newcommand{\QJSD}{\mathcal{J}_Q}

\begin{document}

\title{\textbf{Quantum correlations and Basis-Independent Coherence Distribution
in Two Gravitational Cat States}}

\author{Mostafa Mansour \orcidlink{0000-0003-0821-0582}}\email{mostafa.mansour.fsac@gmail.com}
\affiliation{LMHEP, Department of Physics, Faculty of Sciences Ain Chock,\\ Hassan II University, Casablanca, Morocco}

\author{Mansoura Oumennana }\email{oum.mans@gmail.com}
\affiliation{LMHEP, Department of Physics, Faculty of Sciences Ain Chock,\\ Hassan II University, Casablanca, Morocco}

\begin{abstract}
We study the distribution of quantum correlations and basis-independent coherence in a pair of massive particles confined in a double-well
potential and coupled through their mutual Newtonian gravitational
interaction. Non-classical correlations are characterized using Bures distance of entanglement and quantum discord, while coherence is quantified through the square root of the quantum Jensen--Shannon divergence (QJSD) from the maximally mixed state, yielding a measure that is invariant under
arbitrary unitary transformations and is therefore genuinely basis-independent. The total coherence $C_T$ decomposes into two operationally distinct contributions: the \emph{collective} coherence $\CC$, which captures quantum correlations between the two subsystems, and the \emph{localized} coherence $\CL$, which captures the intrinsic quantum coherence of each individual subsystem.
 We analyze how temperature $T$, the gravitational coupling $\Delta$, and the single-particle energy scale $w$
govern the redistribution of coherence between its collective and
localized components. Our results show that $\CL$ is more robust
against thermal fluctuations than $\CC$, and that increasing $\Delta$
preferentially enhances collective coherence by strengthening
gravitationally induced inter-particle correlations. 
\vspace{10pt}

\end{abstract}

\maketitle
\newpage
\setcounter{page}{1}

\section{Introduction}
\label{sec:intro}

Unifying general relativity with quantum mechanics remains a fundamental open problem in theoretical physics.
\cite{Kiefer2007,Rovelli1998,Mukhi2011}. Beyond the ambitious quest for
a complete theory of quantum gravity, a more immediate and experimentally
accessible question has attracted growing attention: can one design a
laboratory experiment that unambiguously demonstrates whether the
gravitational interaction has an intrinsically quantum nature? In a seminal
proposal, Bose \textit{et al.}\ \cite{Bose2017} and Marletto \& Vedral
\cite{Marletto2017} independently argued that the detection of
entanglement between two massive particles interacting exclusively
through gravity would constitute sufficient evidence for the quantum
character of the gravitational field, since a classical mediator constrained
to local operations and classical communication (LOCC) cannot generate
entanglement.

A particularly clean theoretical model for this scenario was introduced
by Anastopoulos and Hu \cite{Anastopoulos2020}, who studied a pair of massive
particles in a one-dimensional double-well potential coupled through their
mutual Newtonian interaction. When each particle is in a superposition
of the two potential minima, the resulting bipartite quantum state, a
so-called \emph{gravitational cat} (gravcat) state, becomes entangled
through the position-dependent gravitational potential energy. Rojas and
Lobo \cite{Rojas2023} extended this analysis to finite temperature by
coupling the gravcat system to a thermal bath, deriving the thermal
density operator and computing the concurrence and $\ell_1$-norm
of coherence. Their work established how the threshold temperature for
entanglement survival depends on the gravitational coupling and the
energy scale of the individual particles, providing quantitative
benchmarks for future experiments. Furthemore, Lobo \textit{et al.} \cite{Lobo_LIV} proposed a Lindblad equation derived from a dispersion relation that takes into account corrections arising from quantum gravity. They then compared the concurrence of the gravcat model in the case of standard evolution, as well as in two cases of decoherence induced by Lorentz Invariance Violation (LIV). In another study, the authors explored how the quantum nature of spacetime and its associated deformed symmetries influence the nonclassical correlations and the quantum coherence of a two-particle system \cite{Lobo_spacetime}. Several subsequent works have considerably enriched the gravcat literature.
Rahman~et~al.~\cite{rahman} investigated the resource character of gravcat states under thermal,
stochastic, and power-law noise, establishing a robustness hierarchy in which
steerability and Bell nonlocality are more fragile than entanglement, which in turn is
more fragile than coherence.
Haddadi~et~al.~\cite{saeed} showed that classical correlations in a dephasing channel can
partially shield quantum coherence and local quantum Fisher information in the gravcat
system.
The thermal hierarchy of quantum features has also been studied in Ref.~\cite{r3}. The influence of gravitational and thermal effects on skew information correlations and local quantum Fisher information in two gravitational cat states was studied in \cite{dahbi}. The present work complements these studies by providing an observer-independent
decomposition of the total coherence, distinguishing the collectively generated
component from the locally stored one.

Quantum coherence is a central resource in quantum information
processing \cite{Streltsov2017,Baumgratz2014,Tan2016,MO_Coh,MB_Coh, Chouiba2026}, playing a key
role in quantum computation \cite{Ahnefeld2022}, metrology
\cite{Giovannetti2004}, and thermodynamics \cite{Lostaglio2015}. The resource theory of coherence, initiated by Baumgratz, Cramer,
and Plenio \cite{Baumgratz2014}, defines coherence relative to a
preferred incoherent basis, making it an inherently \emph{basis-dependent}
quantity. While this is natural in many contexts where a physical
observable singles out a preferred reference frame, it also introduces conceptual ambiguity: the same quantum state can be considered
coherent or incoherent depending on the choice of basis.

To overcome this limitation, Radhakrishnan \textit{et al.}\
\cite{Radhakrishnan2016,Radhakrishnan2019} proposed a
\emph{basis-independent} coherence measure obtained by
replacing the set of incoherent states with the unique maximally mixed
state $\rhoI = \mathbb{1}/d$ as the reference state. Since $\rhoI$ is
invariant under all unitary transformations, the resulting coherence
measure $\CT(\rho) = \mathcal{D}(\rho,\rhoI)$ is inherent to the quantum state, independent of any choice of basis. Here
$\mathcal{D}$ denotes the square root of the QJSD, which satisfies the
triangle inequality and defines a proper metric in the density matrix space \cite{Endres2003,Lamberti2008}. 

Furthermore, the total coherence admits contributions from collective coherence $\mathcal{C}_C$, which measures the contribution of inter-subsystem correlations, and localized coherence $\mathcal{C}_L$, which captures the coherence intrinsic to each individual subsystem. This decomposition requires only the reduced density matrices, thereby avoids any optimization over separable states, and has been applied to spin chains \cite{Karpat2014,Malvezzi2016}, the transverse Ising model, GHZ and W states \cite{Radhakrishnan2019}, and has more recently been extended to relativistic settings such as de Sitter spacetime \cite{Elghaayda2026}, as well as to condensed-matter platforms such as dipolar coupled electrons in double quantum-dot molecules \cite{Bouafia2026}.

In this work, we apply the basis-independent coherence framework to the
gravcat model of Rojas and Lobo \cite{Rojas2023}. Our central goal is to
go beyond the $\ell_1$-norm studied in Ref.~\cite{Rojas2023} and provide
an observer-independent picture of how quantum coherence is generated,
distributed, and degraded by the interplay of gravitational coupling and
thermal fluctuations. Specifically, we address the following questions:
In what proportion is the coherence stored collectively
(gravitationally induced correlations) versus locally (single-particle
superpositions)? and does the threshold temperature for the survival of collective coherence coincide with the entanglement threshold, or does it reveal a distinct energy scale?

The paper is organized into the following sections: In Sec.~\ref{sec:model} we present
the gravcat Hamiltonian, the thermal density operator, and we
introduce the reduced states needed for the coherence decomposition. In
Sec.~\ref{sec2} we review the basis-independent coherence
framework and define the quantities $\CT$, $\CC$, and $\CL$ explicitly
for the gravcat system. Section \ref{sec:results} presents our numerical
results and physical discussion. We conclude in Sec.~\ref{sec:concl}.
Throughout, we use units $\hbar = k_B = 1$ unless otherwise stated.

\section{Setup}\label{sec:model}
\subsection{Hamiltonian and eigensystem}

We consider two identical massive particles, each of mass $m$, confined
in an even one-dimensional double-well potential whose two minima are
distanced by $L$. The two localized eigenstates of each
particle are denoted $\ket{\pm}$, satisfying $\hat{x}\ket{\pm}=\pm
L/2\,\ket{\pm}$. Following the Landau--Lifshitz approximation
\cite{Anastopoulos2020}, these position eigenstates can be expressed using the ground state $\ket{0}$ and first excited state $\ket{1}$
of the double-well as
\begin{equation}
  \ket{\pm} = \frac{1}{\sqrt{2}}\bigl(\ket{0} \pm \ket{1}\bigr).
  \label{eq:localized_states}
\end{equation}
The two-level truncation is justified by the large energy gap to higher
excited states in the Wentzel-Kramers-Brillouin (WKB) approximation \cite{Anastopoulos2020,Rojas2023}.

\begin{figure}[ht]
\centering
\includegraphics[scale=0.35]{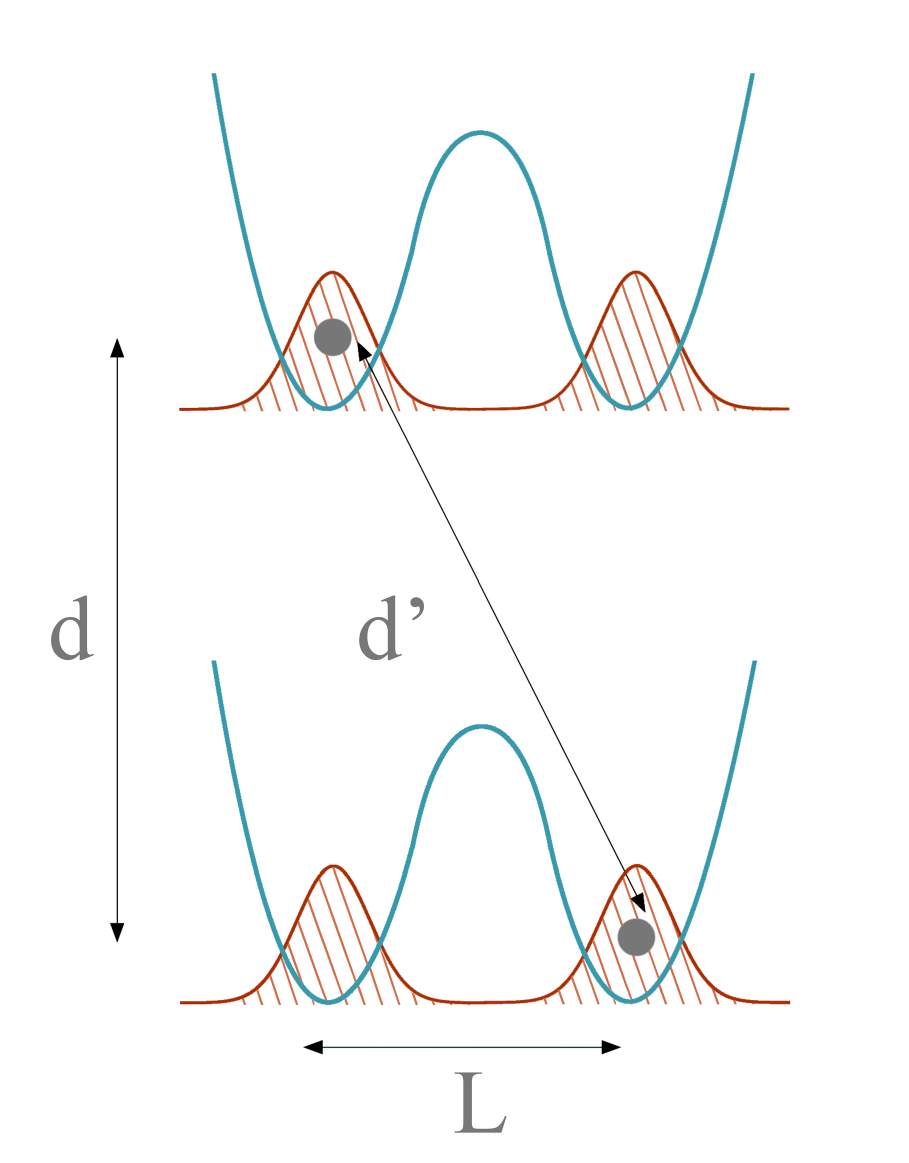}
\caption{Illustration of the gravcat model. Two particles are confined in a double-well potential with minima distanced by $L$. The parameter $d$ $(d')$ denotes the separation between the particles when they occupy the same (opposite) relative minima.}
\label{fig:GCat}
\end{figure}

The two particles are placed at a center-to-center separation $d$ when
occupying the same potential minimum, or $d'$ when at opposite minima, as depicted in Figure~\ref{fig:GCat}. 

Their mutual Newtonian gravitational interaction couples
the different spatial degrees of freedom. In the two-level approximation, the full
bipartite Hamiltonian reads \cite{Anastopoulos2020,Rojas2023}
\begin{equation}
  \mathcal{H} = \frac{w}{2}\bigl(\sigma_z \otimes \mathbb{1}
  + \mathbb{1} \otimes \sigma_z\bigr)
  - \Delta\,\bigl(\sigma_x \otimes \sigma_x\bigr),
  \label{eq:hamiltonian}
\end{equation}
with $\sigma_{x,z}$ being the Pauli operators, $w$ is the energy
splitting between the ground and first excited states of the individual
double-well (setting the single-particle energy scale), and $\Delta$
parametrizes the gravitational coupling strength:
\begin{equation}
  \Delta = \frac{\alpha}{2}\left(\frac{1}{d} - \frac{1}{d'}\right),
  \qquad \alpha = Gm^2,
  \label{eq:Delta}
\end{equation}
with $G$ the universal gravitational constant. The parameter $\Delta$
encodes the anisotropy of the gravitational potential energy over the
four branches of the spatial superposition.

The Hamiltonian \eqref{eq:hamiltonian} is diagonalized in the standard
two-qubit basis $\{\ket{00},\ket{01},\ket{10},\ket{11}\}$ to yield four
eigenvalues:
\begin{equation}
  \varepsilon_1 = -\Delta, \quad
  \varepsilon_2 = \sqrt{\Delta^2 + w^2}, \quad
  \varepsilon_3 = -\sqrt{\Delta^2 + w^2}, \quad
  \varepsilon_4 = \Delta,
  \label{eq:eigenvalues}
\end{equation}
with the corresponding eigenstates 
\begin{align}
  \ket{\varphi_1} &= \frac{1}{\sqrt{2}}\bigl(\ket{01} + \ket{10}\bigr),
  \label{eq:phi1}\\
  \ket{\varphi_2} &=\frac{1}{\sqrt{1+\xi_{-}^{2}}}\left(\ket{00}+\xi_{-}\ket{11}\right),
  \label{eq:phi2}\\
  \ket{\varphi_3} &= \frac{1}{\sqrt{1+\xi_{+}^{2}}}\left(\ket{00}+\xi_{+}\ket{11}\right),
  \label{eq:phi3}\\
  \ket{\varphi_4} &= \frac{1}{\sqrt{2}}\bigl(\ket{01}-\ket{10} \bigr),
  \label{eq:phi4}
\end{align}

where $\xi_{\pm}=\dfrac{\Delta}{w\pm \sqrt{\Delta^{2}+w^{2}}}$.

\subsection{Thermal density operator}
\label{subsec:thermal_rho}

The system is placed in thermal equilibrium with a reservoir characterized by $\beta = 1/(k_B T)$. The thermal state is
\begin{equation}
  \rho(T) = \frac{e^{-\beta\mathcal{H}}}{Z},
  \qquad Z = \mathrm{Tr}\!\left[e^{-\beta\mathcal{H}}\right]
  = \sum_{k=1}^{4} e^{-\beta\varepsilon_k},
  \label{eq:thermal_state}
\end{equation}
Due to the symmetry of the Hamiltonian \eqref{eq:hamiltonian} under
exchange of the two particles and under the $\mathbb{Z}_2$ parity
$\sigma_z \to -\sigma_z$, the density matrix $\rho(T)$ takes the
\emph{X-shape} in the computational basis
$\{\ket{00},\ket{01},\ket{10},\ket{11}\}$:
\begin{equation}
  \rho(T) =
  \begin{pmatrix}
    \rho_{11} & 0 & 0 & \rho_{14} \\
    0 & \rho_{22} & \rho_{23} & 0 \\
    0 & \rho_{23} & \rho_{22} & 0 \\
    \rho_{14} & 0 & 0 & \rho_{44}
  \end{pmatrix},
  \label{eq:density_matrix}
\end{equation}
where all matrix elements are real, the symmetry $\rho_{22}=\rho_{33}$
follows from the exchange symmetry of the two identical particles. The
explicit expressions for the non-zero elements are \cite{Rojas2023}:
\begin{align}
  \rho_{11} &= \frac{1}{Z}\left[
    e^{-\beta\varepsilon_2}\frac{\xi_{-}^{2}}{1+\xi_{-}^{2}}-
    + e^{-\beta\varepsilon_3}\frac{\xi_{+}^{2}}{1+\xi_{+}^{2}}
  \right],
  \label{eq:rho11}\\
  \rho_{14} &= \frac{1}{Z}\left[
    e^{-\beta\varepsilon_2}\frac{\xi_{-}}{1+\xi_{-}^{2}}
    + e^{-\beta\varepsilon_3}\frac{\xi_{+}}{1+\xi_{+}^{2}}
  \right],
  \label{eq:rho14}\\
  \rho_{22} &= \frac{1}{Z}\left[
    \frac{e^{-\beta\varepsilon_1} + e^{-\beta\varepsilon_4}}{2}
  \right],
  \label{eq:rho22}\\
  \rho_{23} &= \frac{1}{Z}\left[
    \frac{e^{-\beta\varepsilon_1} - e^{-\beta\varepsilon_4}}{2}
  \right],
  \label{eq:rho23}\\
  \rho_{44} &= \frac{1}{Z}\left[
    e^{-\beta\varepsilon_2}\frac{1}{1+\xi_{-}^{2}}
    + e^{-\beta\varepsilon_3}\frac{1}{1+\xi_{+}^{2}}
  \right].
  \label{eq:rho44}
\end{align}
One can readily verify that $\mathrm{Tr}[\rho(T)] = \rho_{11} + 2\rho_{22} +
\rho_{44} = 1$.

\subsection{Reduced density matrices and product state}
\label{subsec:reduced}

The reduced density matrix of subsystem $A$ is obtained by taking the partial trace over subsystem $B$:
\begin{equation}
  \rho_A = \mathrm{Tr}_B[\rho(T)]
  = \begin{pmatrix}
    \rho_{11} + \rho_{22} & 0 \\
    0 & \rho_{22} + \rho_{44}
  \end{pmatrix}.
  \label{eq:rho_A}
\end{equation}
By the exchange symmetry of $\mathcal{H}$, we have $\rho_A = \rho_B
\equiv \rho_1$. This diagonal form of $\rho_1$ results straightforwardly from the X-structure of the overall state and simplifies the subsequent coherence calculations considerably. We define
\begin{equation}
  p \equiv \rho_{11} + \rho_{22}, \qquad 1-p = \rho_{22} + \rho_{44},
  \label{eq:p}
\end{equation}
so that $\rho_1 = \mathrm{diag}(p, 1-p)$.

The \emph{product state} $\piro$, which is the closest product state to
$\rho$ under the QJSD metric \cite{Radhakrishnan2019}, is given by the
tensor product of the reduced density matrices:
\begin{equation}
  \piro = \rho_1 \otimes \rho_2 = \rho_1 \otimes \rho_1
  = \begin{pmatrix}
    p^2 & 0 & 0 & 0 \\
    0 & (1-p)p & 0 & 0 \\
    0 & 0 & (1-p)p & 0 \\
    0 & 0 & 0 & (1-p)^2
  \end{pmatrix}.
  \label{eq:product_state}
\end{equation}
Note that $\piro$ is diagonal in the computational basis: it carries no
coherence whatsoever. Its entropy is $S(\piro) = 2S(\rho_1)$, where
$S(\rho_1) = -p\log_2 p - (1-p)\log_2(1-p)$ is the binary Shannon
entropy of the marginal.

\section{Entanglement, Quantum Discord and Basis-Independent Coherence Measures}\label{sec2}

\subsection{Bures distance entanglement}

A variety of entanglement quantifiers have been introduced for composite systems. Some are distance-based measures. Other measures are convex roof measures. Among distance-based
measures, the one employed in this work is the Bures distance \cite{bu04, bu1, bu2, bu3, bu4, bu5, bu6}
\begin{equation}\label{bures}
B(\rho)= \sqrt{2 -2\sqrt{\mathcal{F}_{sep}(\rho)}},
\end{equation}

where $\mathcal{F}_{sep}(\rho)$ is the fidelity of separability defined as the maximum of Ulhmann's fidelity \cite{bu7} $\mathcal{F}(\rho, \sigma)=\left(\text{Tr}\left(\sqrt{\sqrt{\rho}\sigma\sqrt{\rho}}\right)\right)^{2}$, with $\sigma$ being a state from the set of separable states. For two-qubit states, the fidelity of separability can therefore be given as a function of concurrence \cite{C_Wootters, Wootters1998, Hill1997} as \cite{bu04, Wei2003}
\begin{equation}
\mathcal{F}_{sep}(\rho)=\frac{1}{2}\left(1+\sqrt{1-C\left(\rho\right)^{2}}\right).\label{eq:Fs-2}
\end{equation}

Finally, Bures distance can be rewritten in terms of the concurrence. For two qubit states, Bures distance $B(\rho)$ is zero for separable states $(C(\rho)=0)$ and it is equal to $(\sqrt{2-\sqrt{2}})$ for maximally entangled states. Here, we adopt the normalized Bures distance expresse as
\begin{equation}\label{bures01}
B(\rho)=\frac{1}{\sqrt{2-\sqrt{2}}} \left[\sqrt{2 -\sqrt{2+ 2\sqrt{1-C(\rho)^2}}} \right],
\end{equation}
thereby ensuring that $(B(\rho)=1)$ (\ref{bures01}) for maximally entangled states. 

We recall that for a state with the X-form concurrence has a simple closed formula, particularly for the thermal density matrix (\ref{eq:density_matrix}) it is given by  
\begin{equation}
C = \max\{0,\, C_1,\, C_2\},
\end{equation}
where
\begin{equation}
C_1 = 2\bigl(|\rho_{23}| - \sqrt{\rho_{11}\,\rho_{44}}\bigr), \qquad
C_2 = 2\bigl(|\rho_{14}| - \rho_{22}\bigr).
\end{equation}

\subsection{Quantum discord}
Quantum discord (QD) \cite{discord} quantifies quantum correlations beyond entanglement in a composite system. For a two-qubit system $AB$, QD is defined as 
\begin{equation}
\label{eq17}
\text{QD}(\rho_{AB})=  {\mathcal I}(\rho_{AB})-{\mathcal C}{\mathcal C}(\rho_{AB}),
\end{equation}
and ${\mathcal I}(\rho_{AB})$ is the quantum mutual information  
\begin{equation}
\label{eq18}
{\mathcal I}(\rho_{AB})=-{\mathcal S}(\rho_{AB})+{\mathcal S}(\rho_A)+{\mathcal S}(\rho_B),
\end{equation}
and ${\mathcal C}{\mathcal C}(\rho_{AB})$ are the classical correlations that can be defined as \cite{HendersonL}
\begin{equation}
\label{eq19}
{\mathcal C}{\mathcal C}(\rho_{AB})={\mathcal S}(\rho_A)- \min_{\{\pi_{i}^{B}\}}\sum_{i}p_{i}{\mathcal S}(\rho_{A|i}).
\end{equation}
Here $ {\mathcal S}(\rho)=-\text{Tr}(\rho \log_2 \rho) $ is the von Neumann entropy. $\/{\left\{ \pi_{B}^{i}\right\}}={|i_{B} \rangle \langle i_{B} |} $  denotes a complete set  of orthonormal projective measurements that act only on subsystem $B$. The conditional state of the subsystem $A$, corresponding to the measurement outcome $i$ on $B$ is 
 $\rho_{A|i}=\frac{\text{Tr}_{B}(\pi_{B}^{i}\rho_{AB} \pi_{B}^{i})}{p_{i}}$, where $p_{i}= \text{Tr}_{AB} ( \pi_B^i \rho_{AB} \pi_B^i)  $  represents the probability of obtaining the outcome $i$.

Using Eqs. \eqref{eq18}  and \eqref{eq19}, the quantum discord \eqref{eq17} can be expressed as follows \cite{HendersonL}
\begin{equation}
 \label{eq2021}
\text{QD}(\rho_{AB}) = \min_{               \left\lbrace \pi_B^i      \right\rbrace        }[{\mathcal S}(\rho_{AB}/{\left\{ \pi_{B}^{i}\right\}} ]+{\mathcal S}(\rho_B)  -{\mathcal S}(\rho_{AB}).
 \end{equation}

Analytical expressions for QD are generally difficult to obtain for general quantum states because they involve an optimization over the conditional entropy. Exact closed-form expressions have been derived only for a limited class of states, such as X-states.  In the present work, the thermal state of our system takes precisely the X-state form given in (\ref{eq:density_matrix}). For such states, QD is redefined by the following expression \cite{discx,discxa}

\begin{equation}
\label{eq22}
\text{QD}(\rho_{AB})=\min\{\mathcal{Q}_1,\mathcal{Q}_2\},
\end{equation}
with  $$ \mathcal{Q}_i=H(\rho_{11 }+\rho_{33 }) +\sum_{k=1}^{4} \lambda_k \log_2(\lambda_k)+{\Lambda}_{i}.$$

$ { \Lambda}_1=H(\delta)$ and ${ \Lambda}_2=-\sum_{n=1}^{4}\rho_{nn} \log_2(\rho_{nn}) -H(\rho_{11 }+\rho_{33})$.\\
With $ \delta =\frac{\sqrt{[1-2(\rho_{33 }+\rho_{44 })]^2+4 (|\rho_{14}|+|\rho_{23}|)^2}}{2}+\frac{1}{2}$ and $ H(y)=-y \log_2(y)-(1-y)\log_2(1-y)$ is the binary Shannon entropy.  $\lambda_k$'s denote the eigenvalues of $\rho_{AB}$.

\subsection{The quantum Jensen--Shannon divergence}
\label{subsec:qjsd}

The QJSD between two density
matrices $\rho$ and $\sigma$ is defined as \cite{Brietharre2009,Majtey2005}
\begin{equation}
  \QJSD(\rho,\sigma)
  = S\!\left(\frac{\rho+\sigma}{2}\right)
    - \frac{S(\rho) + S(\sigma)}{2},
  \label{eq:QJSD}
\end{equation}
with $S(\rho)$ being the von Neumann entropy.
The QJSD is symmetric, non-negative, and bounded $(0 \le \QJSD(\rho,\sigma)
\le 1)$, but does not in general satisfy the triangle inequality. Its square root,
\begin{equation}
  \mathcal{D}(\rho,\sigma) = \sqrt{\QJSD(\rho,\sigma)},
  \label{eq:D_metric}
\end{equation}
satisfies the triangle inequality for pure states and is conjectured, and numerically verified, to do so for mixed states as well
\cite{Lamberti2008}. 

\subsection{Total basis-independent coherence}
\label{subsec:CT}

Following Radhakrishnan \textit{et al.}\ \cite{Radhakrishnan2016,Radhakrishnan2019},
the \emph{total} basis-independent coherence of a $d$-dimensional quantum
state $\rho$ is defined as its distance from the unique maximally mixed
(maximally incoherent) state $\rhoI = \mathbb{1}_d/d$:
\begin{equation}
  \CT(\rho) \equiv \mathcal{D}(\rho,\rhoI)
  = \sqrt{S\!\left(\frac{\rho + \mathbb{1}_d/d}{2}\right)
    - \frac{\log_2 d+S(\rho) }{2}}.
  \label{eq:CT}
\end{equation}
For the gravcat system, $d=4$ (two qubits), so $\rhoI = \mathbb{1}_4/4$
and $\log_2 d = 2$. For the thermal gravcat state \eqref{eq:density_matrix}, the matrix
$(\rho + \mathbb{1}_4/4)/2$ retains the X-form and its four eigenvalues
$\lambda_k$ can be obtained analytically, yielding
\begin{equation}
  \CT(\rho) = \sqrt{
    -\sum_{k=1}^{4}\lambda_k \log_2\lambda_k
    - \frac{S(\rho) + 2}{2}
  },
  \label{eq:CT_gravcat}
\end{equation}
where $S(\rho) = -\sum_k e_k \log_2 e_k$ with $e_k = e^{-\beta\varepsilon_k}/Z$
the Boltzmann weights, and the eigenvalues $\lambda_k$ are those of
$(\rho+\mathbb{1}_4/4)/2$.

\subsection{Collective coherence}
\label{subsec:CC}

The collective coherence quantifies the contribution to the total
coherence that arises from inter-subsystem correlations; both quantum
(entanglement) and classical captured by the distance between the full
state $\rho$ and its product-state approximation $\piro$:
\begin{equation}
  \CC(\rho) \equiv \mathcal{D}(\rho,\piro)
  = \sqrt{S\!\left(\frac{\rho + \piro}{2}\right)
    - \frac{S(\piro)+ S(\rho)  }{2}}.
  \label{eq:CC}
\end{equation}
For the gravcat state, $S(\piro) = 2S(\rho_1) = -2[p\log_2 p
+ (1-p)\log_2(1-p)]$. The matrix $(\rho + \piro)/2$ also retains the
X-form, since $\piro$ is diagonal and adds to the diagonal blocks of
$\rho$. Its non-zero entries are:
\begin{align}
  \left(\frac{\rho+\piro}{2}\right)_{11} &= \frac{\rho_{11}+p^2}{2},
  \nonumber\\
  \left(\frac{\rho+\piro}{2}\right)_{22} &= \frac{\rho_{22}+p(1-p)}{2},
  \nonumber\\
  \left(\frac{\rho+\piro}{2}\right)_{14} &= \frac{\rho_{14}}{2},
  \nonumber\\
  \left(\frac{\rho+\piro}{2}\right)_{23} &= \frac{\rho_{23}}{2},
  \nonumber\\
  \left(\frac{\rho+\piro}{2}\right)_{44} &= \frac{\rho_{44}+(1-p)^2}{2}.
  \label{eq:rho_plus_pi_over2}
\end{align}
The four eigenvalues of \eqref{eq:rho_plus_pi_over2} can be obtained
analytically from the $2\times2$ blocks, and thus $\CC(\rho)$ is fully
determined by the parameters $\{T, \Delta, w\}$. Physically, $\CC$ measures the quantum resources encoded in the
correlations between the two gravcat particles. 

\subsection{Localized coherence}
\label{subsec:CL}

The \emph{localized} coherence measures the coherence intrinsic to the
individual subsystems, isolated from any inter-particle correlations. It
is introduced as a measure of the distance between the product state $\piro$ and the
maximally mixed state $\rhoI$:
\begin{equation}
  \CL(\rho) \equiv \mathcal{D}(\piro,\rhoI)
  = \sqrt{S\!\left(\frac{\piro + \mathbb{1}_4/4}{2}\right)
    - \frac{S(\piro) + 2}{2}}.
  \label{eq:CL}
\end{equation}
Since $\piro$ is diagonal [Eq.~\eqref{eq:product_state}], the matrix
$(\piro + \mathbb{1}_4/4)/2$ is also diagonal with entries
\begin{equation}
  \frac{1}{2}\,\mathrm{diag}
    \!\left(p^2+\tfrac{1}{4},\; p(1-p)+\tfrac{1}{4},\;
    p(1-p)+\tfrac{1}{4},\; (1-p)^2+\tfrac{1}{4}\right).
  \label{eq:pi_plus_I_over2}
\end{equation}
The entropy of this matrix is therefore trivially computable:
\begin{align*}
S\!\left(\frac{\pi_\rho + \mathbb{1}_4/4}{2}\right)
&= -\frac{p^2+\frac{1}{4}}{2}\log_2\!\frac{p^2+\frac{1}{4}}{2} \\
&\quad - \left(p(1-p)+\frac{1}{4}\right)\log_2\frac{p(1-p)+\frac{1}{4}}{2} \\
&\quad - \frac{(1-p)^2+\frac{1}{4}}{2}
    \log_2\!\frac{(1-p)^2+\frac{1}{4}}{2}.
\end{align*}
Note that $\CL$ depends on the temperature $T$ only through $p(T)$, the
single-particle marginal occupation probability. This makes $\CL$ a
direct measure of the local single-particle superposition, decoupled from
inter-particle correlations.

\section{Results and Discussion}\label{sec:results}

\subsection{Dynamics of entanglement in gravcat states}

\begin{figure}[H]
    \centering
    \includegraphics[width=1.0\linewidth]{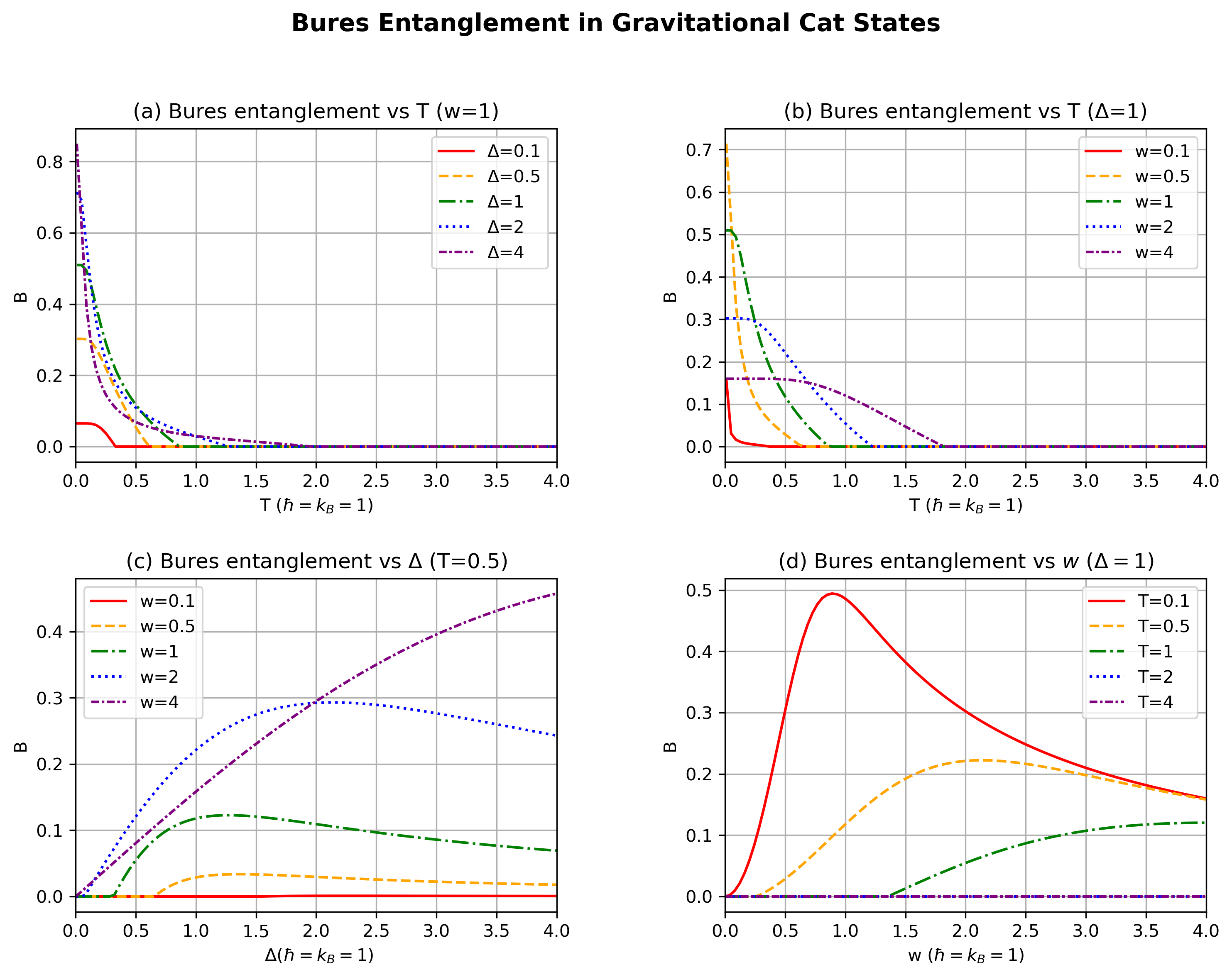}
    \caption{Bures measure of entanglement $B$ as a function of $T$, $\Delta$ and $w$. (a) $B$ vs $T$ for fixed $w=1$ and varying $\Delta$; (b) $B$ vs $T$ for fixed $\Delta=1$ and varying $w$; (c) $B$ vs $\Delta$ for fixed $T=0.5$ and varying $w$; (d) $B$ vs $w$ for fixed $\Delta=1$ and varying $T$.}
    \label{Fig1}
\end{figure}

In Figure~\ref{Fig1}, we plot Bures distance entanglement $B$ as a function of temperature $T$, gravitational coupling $\Delta$, and single-particle energy scale $w$ for the gravcat system. Panel (a) fixes $w=1$ and varies $\Delta$. At $T=0$, Bures distance entanglement increases with $\Delta$, reaching $B\approx 0.9$ for $\Delta=4w$, while remaining nearly zero for $\Delta=0.1w$. This behavior originates from the structure of the ground state: for $\Delta>>w$, the interaction term dominates the Hamiltonian and the ground state tends to a Bell-like state, which manifests higher amounts of entanglement. As temperature rises, Bures distance entanglement decreases monotonically due to thermal mixing, and the threshold temperature for entanglement survival grows with $\Delta$: for $\Delta=4w$ entanglement persists up to $T\approx 2$, whereas for $\Delta=2w$ it vanishes already at $T\approx 1$. Panel (b) fixes $\Delta=1$ and varies $w$. In this panel, the opposite trend is observed: increasing $w$ reduces the near-zero temperature entanglement. Indeed, when $w>>\Delta$, the single-particle energy splitting dominates and the ground state approaches the separable state. Interestingly, however, the temperature at which entanglement disappears increases with $w$. Panel (c) shows Bures distance entanglement as a function of $\Delta$ at fixed $T=0.5$ for several values of $w$. For all values of $w$, the entanglement remains negligible when $\Delta$ is small, indicating that the gravitational coupling is too weak to overcome thermal mixing. As $\Delta$ increases, the entanglement grows rapidly, highlighting the entangling role of the interaction term. However, the dependence is not monotonic: for each value of $w$, Bures measure of entanglement reaches a maximum at a specific gravitational coupling strength $\Delta$, beyond which a gradual decrease is observed. Both the position and the amplitude of this maximum depend on $w$, indicating that the generation of thermal entanglement results from a delicate interplay between the local energy splitting and the interaction strength. Panel (d) presents Bures distance entanglement versus $w$ for several temperatures at fixed $\Delta=1$. At low temperatures, particularly for $T=0.1$, the entanglement exhibits a pronounced maximum near $w\approx \Delta=1$. For very small $w$, the difference of energy between the ground and excited states is small, making the system more susceptible to thermal mixing, which results in the reduced amounts of entanglement. As $w$ slightly increases, the energy gap grows, therefore thermal excitations diminish, which consequently allows for the observed increase of entanglement. However, for $w>>\Delta$, the local energy splitting dominates  the gravitational interaction term, and the ground state tends to a separable product state and thereby reducing the entanglement of the Gibbs state at finite temperatures. For sufficiently high temperatures, thermal mixing dominates the system dynamics and the entanglement is strongly suppressed over the entire range of $w$. These results demonstrate that strong thermal entanglement in the gravcat model requires a balance between the local energy splitting $w$ and the gravitational interaction strength $\Delta$.

\subsection{Dynamics of quantum discord in gravcat states}

\begin{figure}[H]
    \centering
    \includegraphics[width=1.0\linewidth]{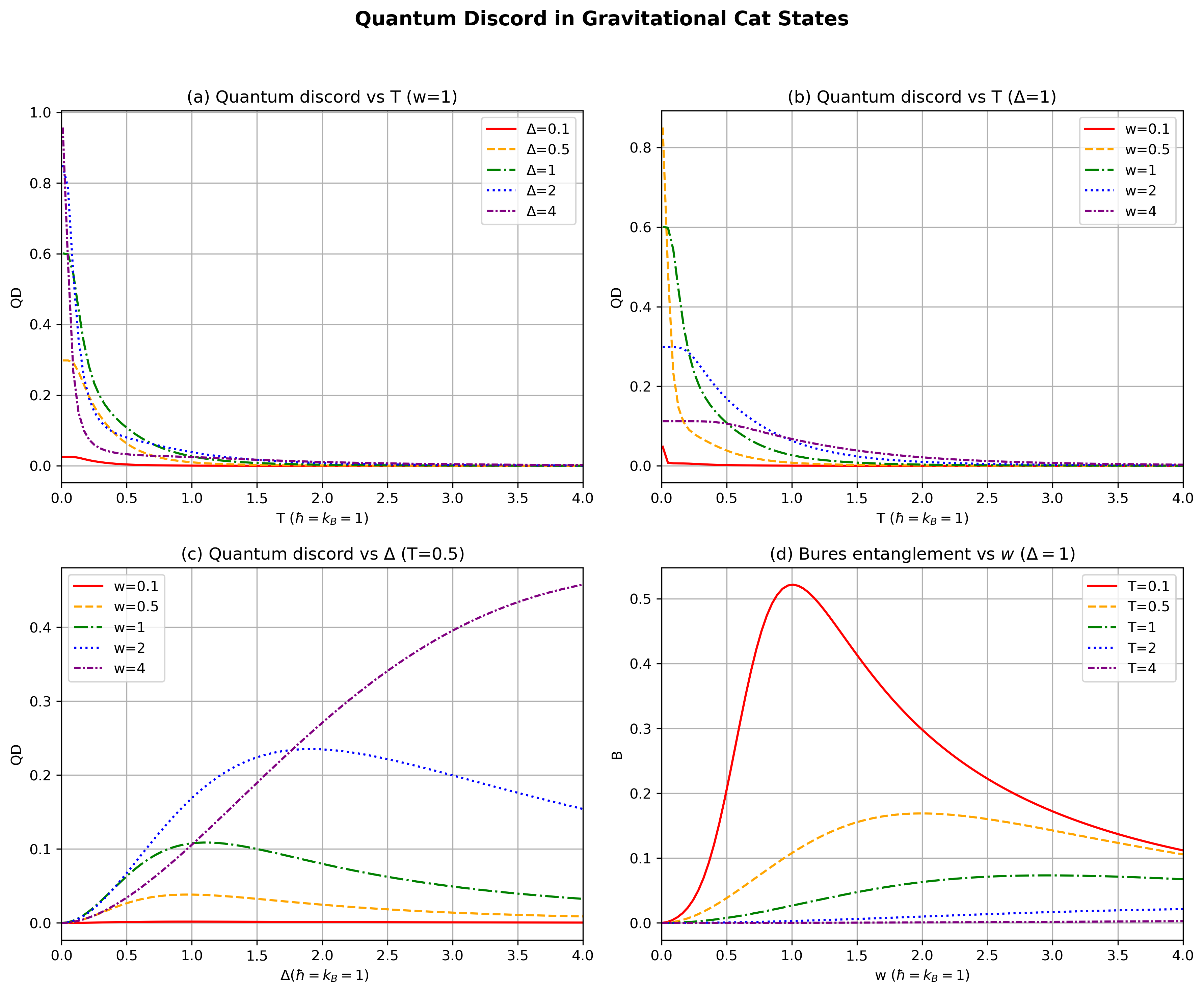}
    \caption{QD  as a function of $T$, $\Delta$ and $w$. (a) QD vs $T$ for fixed $w=1$ and varying $\Delta$; (b) QD vs $T$ for fixed $\Delta=1$ and varying $w$; (c) QD vs $\Delta$ for fixed $T=0.5$ and varying $w$; (d) QD vs $w$ for fixed $\Delta=1$ and varying $T$.}
    \label{Fig2}
\end{figure}

Regarding the dynamics of quantum discord, we observe behavior similar to that of entanglement. A strong gravitational coupling $\Delta$ strengthens  QD for low temperature values. A notable difference, however, is that quantum discord doesn't show the sharp vanishing observed in entanglement at specific temperatures depending on the different values of $\Delta$. Instead, quantum discord keeps decreasing smoothly and monotonically as the temperature increases. On the other hand, Fig.(\ref{Fig2}d) shows that when $w=0$, quantum correlations vanish even if $\Delta \neq 0$. This may appear counterintuitive since the gravitational coupling term is nonzero ($\Delta=1$); however, it can be understood from the structure of the Hamiltonian.  In this system of two gravitational cat states, quantum correlations originate from the non-commutativity between the local field and interaction terms, particularly if we calculate the commutator of the two terms of the Hamiltonian, we find that it is expressed as $-iw\Delta(\sigma_y\otimes \sigma_x+\sigma_x\otimes \sigma_y)$. Physically, each parameter plays a distinct role: the local term proportional to $w$ tends to localize the particles along the $z$-direction, while the interaction term proportional to $\Delta$  promotes collective quantum superpositions. The competition between these non-commuting contributions is precisely what generates nontrivial quantum correlations in the system. Moreover, we see that for both entanglement and quantum discord the higher amounts observed are when the values of $\Delta$ and $w$ are comparable.

\subsection{Effect of Temperature on Coherence Distribution}

We first examine how thermal fluctuations affect the distribution of 
basis-independent coherence. Figure~\ref{Fig2} 
displays $C_T$, $C_C$, and $C_L$ as functions of temperature $T$ for fixed 
values of $\Delta$ and $w$.

\begin{figure}[H]
    \centering
    \includegraphics[width=1.0\linewidth]{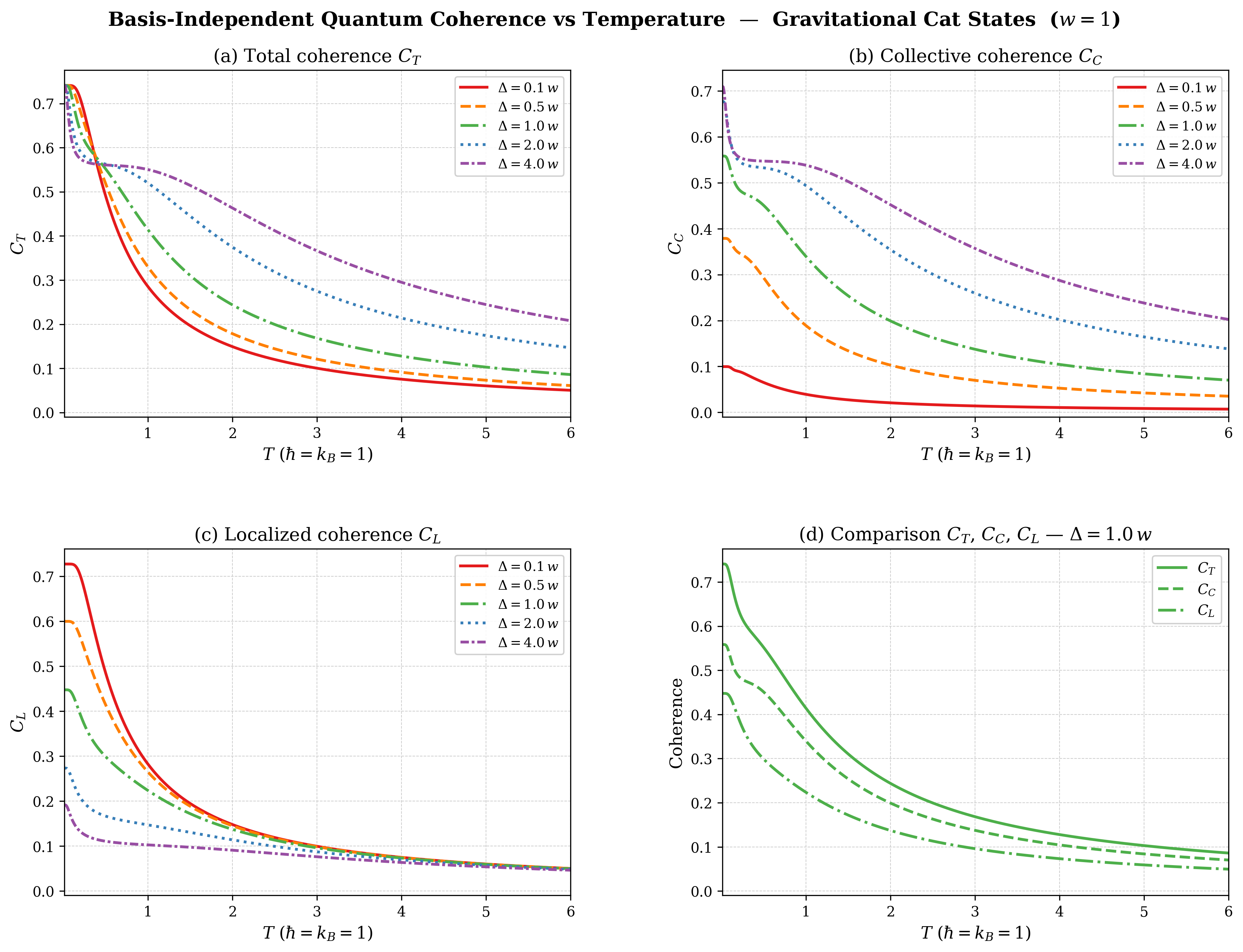}
    \caption{Total, collective and localized coherence vs temperature for $\Delta=0.1w,0.5w,1.0w,2.0w,4.0w$ at fixed $w=1$. (a) $C_T$, (b) $C_C$, (c) $C_L$, (d) comparison for $\Delta=1.0w$.}
    \label{Fig2}
\end{figure}

In Figure~\ref{Fig2}, we present the temperature dependence of total ($C_T$), collective ($C_C$) and localized ($C_L$) basis-independent coherence for five gravitational coupling $\Delta$ values at fixed $w=1$. In panel (a), $C_T$ decreases monotonically with $T$ for all $\Delta$, with a slightly faster decay for smaller $\Delta$ values. Panel (b) shows that $C_C$ also drops with $T$, but its zero-temperature value is smaller than $C_T$ when $\Delta$ is small, and it vanishes around $T\approx2$ for $\Delta=0.1\,w$, whereas larger $\Delta$ values sustain nonzero collective coherence $C_C$ at higher temperatures. Panel (c) reveals that at $T=0$, the localized coherence exhibits larger values for smaller gravitational coupling strength $\Delta$. However, despite these differences in magnitude, $C_L$ follows a similar decay pattern for all values of $\Delta$. This behavior may be attributed to the fact that $w$ is fixed at $1$, and this parameter primarily governs superpositions at the single-particle level. Consequently, keeping $w$ fixed while increasing $\Delta$ enhances the contribution of global superpositions within the system. As a result, a larger fraction of the total coherence $C_T$ is stored in the joint degrees of freedom, \textit{i.e.}, in the correlations between the two particles. Therefore, the share of localized coherence $C_L$ in each independent particle decreases. Panel (d) compares the three coherence measures for $\Delta=1.0\,w$. At low temperatures, $C_T\approx C_C > C_L$, indicating that collective coherence $(C_C)$ contributes slightly more to the total coherence $(C_T)$ of the system than localized coherence $(C_L)$. It is important to emphasize, however, that this behavior is specific to this configuration where $\Delta=w=1$. Increasing $w$ can enhance the contribution of localized coherence as compared to that collective coherence, as will be demonstrated next.

\subsection{Effect of the Single-Particle Energy Scale $w$ on Coherence Distribution}

\begin{figure}[H]
    \centering
    \includegraphics[width=1.0\linewidth]{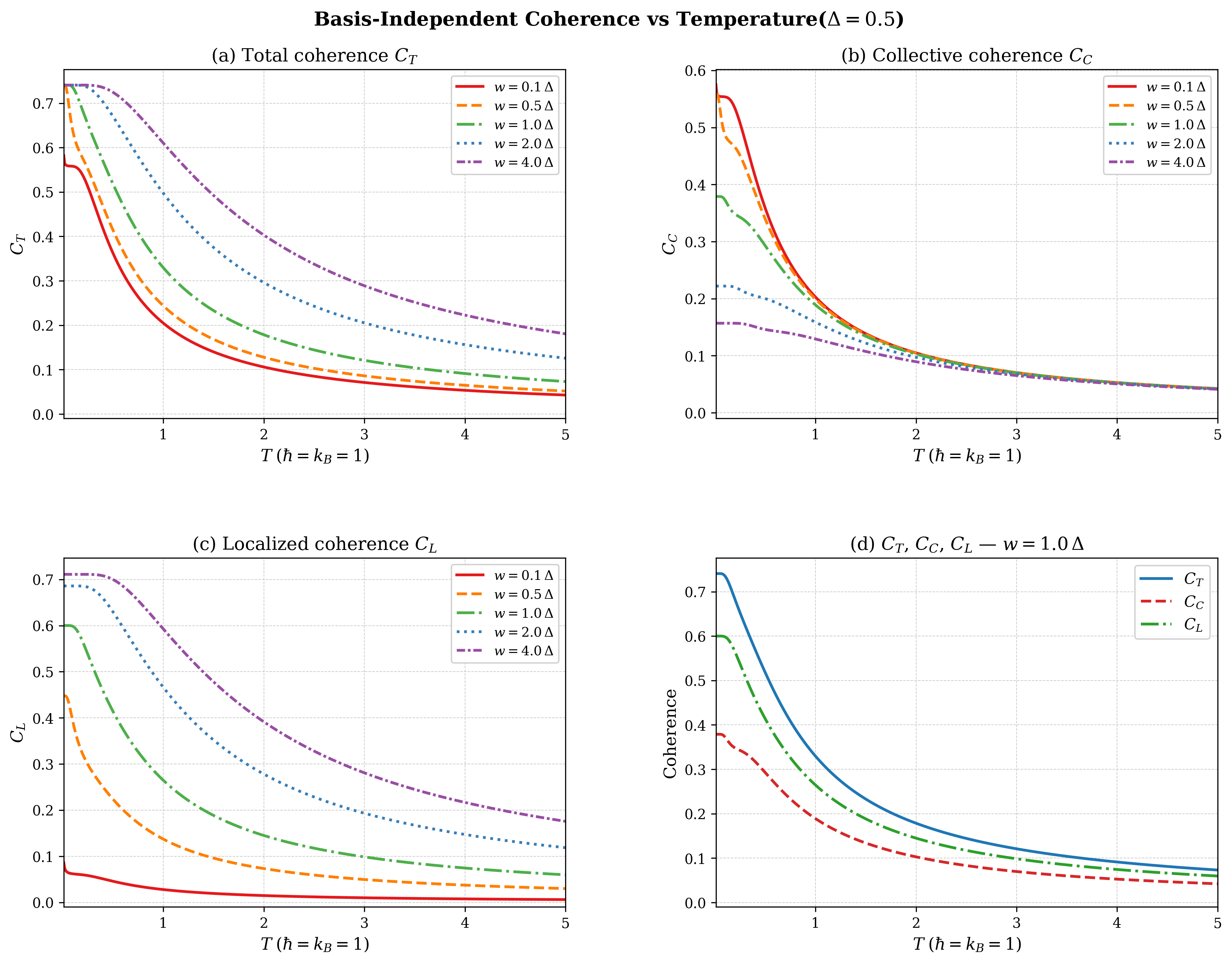}
    \caption{Coherence vs temperature for fixed $\Delta=0.5$ and several $w$. (a) $C_T$, (b) $C_C$, (c) $C_L$, (d) comparison for $w=1$.}
    \label{Fig3}
\end{figure}

We now investigate the influence of the single-particle energy splitting 
$w$, which sets the tunneling scale of each particle in its individual 
double-well potential. Figure~\ref{Fig3} presents $C_T$, $C_C$, and 
$C_L$ as functions of $w$ at fixed $T$ and $\Delta$. 

Figure~\ref{Fig3} displays the temperature dependence of basis-independent coherence for fixed gravitational coupling $\Delta=0.5$ and several values of the single-particle energy splitting $w$. Panel (a) shows the total coherence $C_T$. At $T=0$ we observe that $C_T\approx 0.74$ for all considered $w$ values, except for $w=0.1$, where $C_T\approx 0.58$. As expected, $C_T$ decreases for higher $T$ values due to the incoherent thermal mixing. Nevertheless,  for comparable finite temperatures, larger values of $w$ lead to higher total coherence, indicating an enhanced robustness against thermal effects. Panel (b) presents the collective coherence $C_C$, which exhibits the opposite behavior: at $T=0$, $C_C$ is higher for $w=0.1$ and decreases as $w$ increases. This can be understood from the fact that the gravitational coupling $\Delta$, which is responsible for creating global superpositions between the states of both particles, is fixed at the relatively small value $\Delta=0.5$. Consequently, when $w>\Delta$, single-particle superpositions become more prominent than global superpositions, leading to a reduction in the collective contribution to the total coherence.  Panel (c) shows the dependence of localized coherence $C_L$ on $w$. Localized coherence increases with $w$, both at zero and finite temperatures. Indeed, a larger local energy gap protects the single-particle superposition against thermal fluctuations. Panel (d) compares the three quantities for $w=1.0$ and $\Delta=0.5$. In this configuration the share of localized coherence is the one contributing the most to the total coherence of the gravcat system.

\subsection{Effect of Gravitational Coupling $\Delta$ on Coherence Distribution}

\begin{figure}[H]
    \centering
    \includegraphics[width=1.0\linewidth]{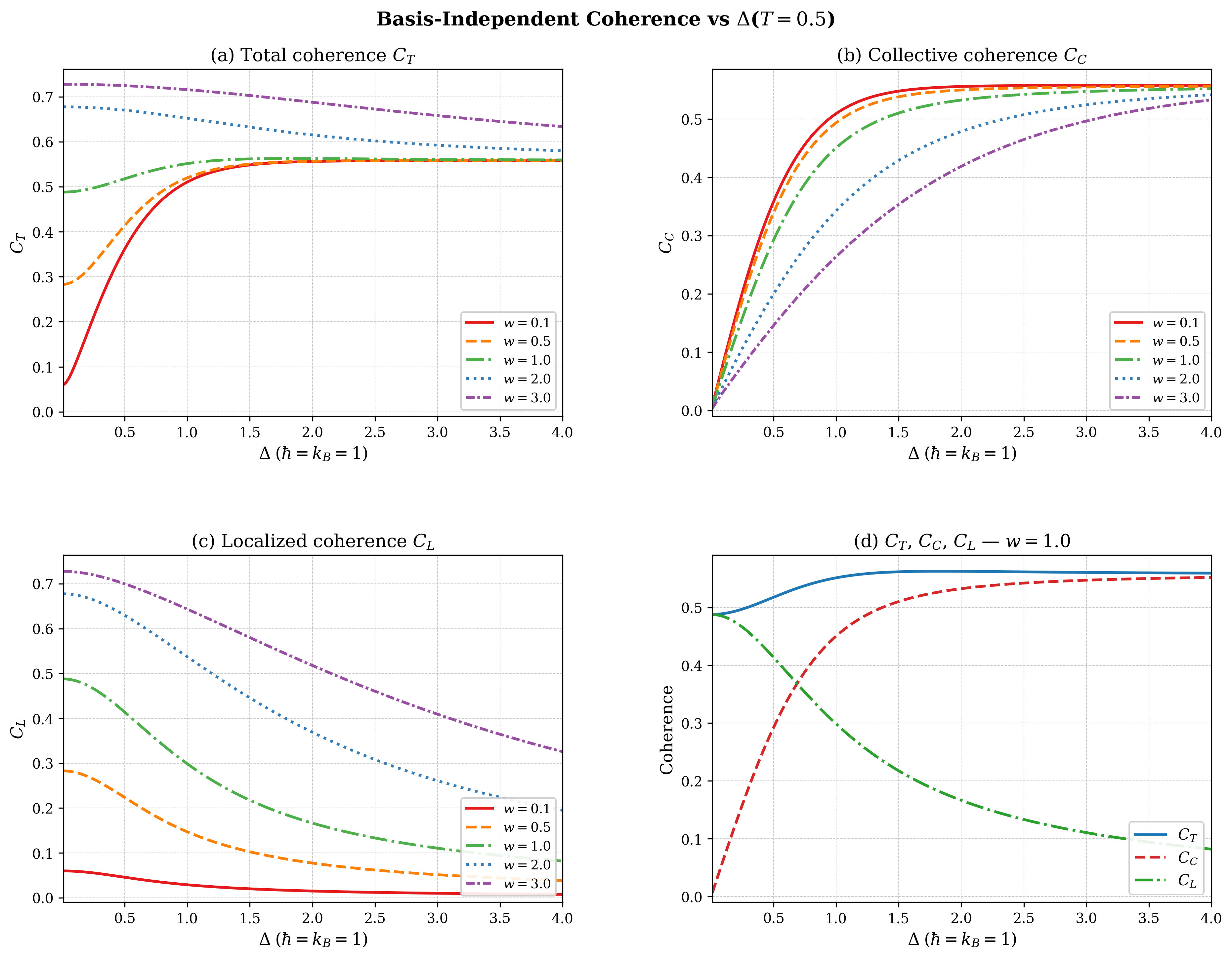}
    \caption{Coherence vs $\Delta$ at fixed $T=0.5$ for several $w$. (a) $C_T$, (b) $C_C$, (c) $C_L$, (d) comparison for $w=1$.}
    \label{Fig4}
\end{figure}

We now examine the role of the gravitational coupling parameter $\Delta$, 
which encodes the anisotropy of the Newtonian interaction energy over the 
four branches of the spatial superposition. Figure~\ref{Fig4} shows 
$C_T$, $C_C$, and $C_L$ as functions of $\Delta$ at fixed $T$ and $w$.

Figure~\ref{Fig4} presents the basis-independent coherence as a function of the gravitational coupling $\Delta$ at fixed temperature $T=0.5$ for a range of values of the single-particle energy splitting $w$. Panel (a) shows the total coherence $C_T$: it increases monotonically with $\Delta$ for all $w$, starting from a finite value at $\Delta=0$ (where only local coherence exists) and saturating at a plateau. Panel (b) displays the collective coherence $C_C$, which grows from zero at $\Delta=0$ (uncoupled qubits) and rises steeply before saturating. The initial slope is larger for smaller $w$, indicating that a small $w$ facilitates the build-up of gravitationally induced correlations. Panel (c) shows the localized coherence $C_L$, which gradually decreases as $\Delta$ increases for each fixed value of $w$. However, for a given $\Delta$, the curves corresponding to larger $w$ consistently exhibit higher values of $C_L$. This behavior confirms that localized coherence is primarily determined by the single-particle marginal states and is therefore governed mainly by the single-particle energy splitting $w$, while being only indirectly affected by the gravitational coupling through the redistribution of coherence between local and collective degrees of freedom. Panel (d) compares $C_T$, $C_C$ and $C_L$ for the representative case $w=1$. In the absence of gravitational coupling ($\Delta=0$), the total coherence coincides with the localized coherence, $C_T=C_L$, while the collective coherence vanishes ($C_C=0$) since no interparticle correlations can be generated. As $\Delta$ increases, $C_L$
gradually decreases, whereas $C_C$ grows due to the increasing contribution of gravitationally induced global superpositions. This trend continues until a crossover occurs around $\Delta \approx 0.7$, beyond which the contribution of gravitationally driven collective coherence exceeds that of purely local single-particle superpositions, becoming the dominant component of the system's total coherence $C_T$.

\subsection{Comparison between Quantum coherence, Quantum discord and Entanglement in gravcat states}

\begin{figure}[H]
    \centering
    \includegraphics[width=1.0\linewidth]{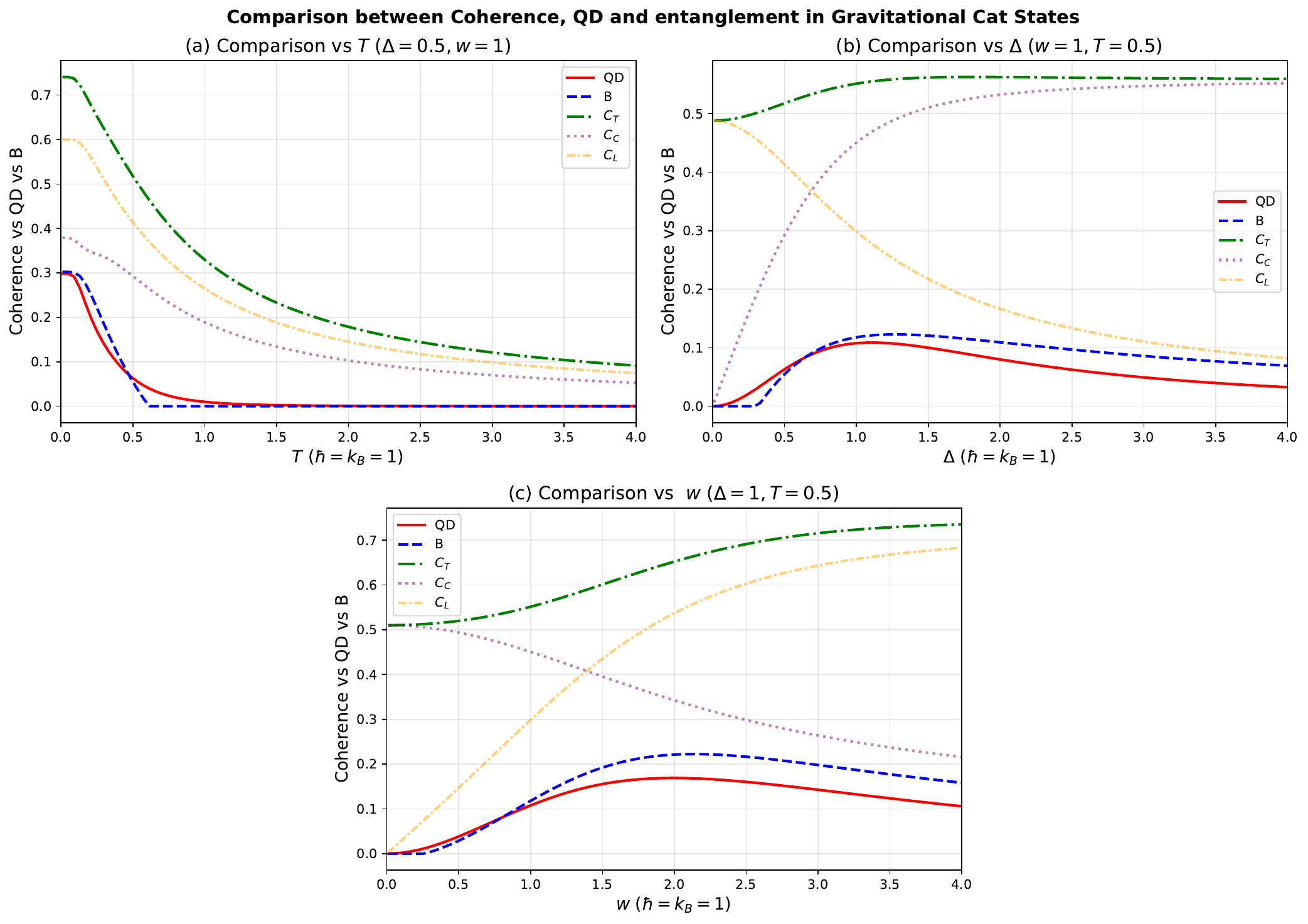}
    \caption{Comparative plots of coherence, QD and Bures measure of entanglement (a) vs  $T$ for $\Delta=0.5, ~ w=1$, (b) vs $\Delta$ for $w=1,~T=0.5$, and (c) vs $w$, for $\Delta=1,~ T=0.5$.}
    \label{Fig5}
\end{figure}

In this part, we aim to understand the trade-off between localized and collective coherence. 
As shown in Fig.(\ref{Fig5}) both collective and localized aspects contribute to the overall coherence of the system. The relative dominance of one over the other depends mainly on the choice of the system parameters. In Fig.(\ref{Fig5}b), $\Delta=0$ corresponds to the absence of coupling between the two qubits. In this case, no quantum correlations are generated between the particles ($B=QD=0$). Nevertheless, the total coherence is non-zero, which is solely due to the presence of local superpositions within each subsystem. As soon $\Delta \neq 0$ collective coherence $C_C$ is generated, and the same is observed for QD.  However, for this configuration where ($T=0.5$), Bures measure of entanglement becomes non-zero $B\neq 0$ only for $\Delta>0.3$. This indicates that entanglement requires stronger gravitational coupling $\Delta$  to overcome the thermal mixing induced by the finite temperature. Therefore, increasing $\Delta$ leads to the redistribution of coherence which becomes less stored in individual qubits and more stored in joint degrees of freedom of the global two-qubit system. On the other hand, Fig.(\ref{Fig5}c) shows that when $w=0$, both entanglement and quantum discord are zero even if the coupling parameter $\Delta \neq 0$. We have explained earlier that quantum correlations arise due to the non-commutativity between the local field and the interaction terms, however; for $w=0$ their commutator vanishes. Nevertheless, the two-qubit system is coherent, as indicated by the total coherence $C_T \approx 0.5$, which arises entirely from collective coherence $C_C$ due to the gravitational coupling term. When $w\neq 0$, there is a competition between the local and interaction terms, leading to the generation of quantum correlations. The effect of $w$ is more visible on coherence: increasing $w$ increases the share of localized coherence and decreases that of the collective coherence stored in the correlations between the two qubits.

\section{Conclusion}\label{sec:concl}

 We investigate the distribution of basis-independent quantum coherence 
in the gravitational cat (gravcat) model, decomposing the total coherence $C_T$ into 
a collective component $C_C$, arising from gravitationally induced inter-particle 
correlations, and a localized component $C_L$, intrinsic to each individual 
particle. This decomposition, based on the quantum Jensen--Shannon divergence metric, 
provides an observer-independent description of how coherence is generated and redistributed 
under the competing influences of gravitational coupling $\Delta$, single-particle 
energy $w$, and temperature $T$. Total coherence decreases monotonically 
with temperature; more importantly, $C_L$ proves substantially more robust against thermal 
noise than $C_C$. This provides a coherence-theoretic interpretation of the well-known 
hierarchy of quantum resources in gravcat states: local superpositions outlast collectively generated correlations as temperature rises. Increasing $\Delta$ selectively amplifies $C_C$, strengthening gravitationally induced inter-particle 
coherence. Moreover, $C_C$ vanishes at a threshold temperature distinct from and lower than that of $C_L$,  confirming that the two components carry independent physical information inaccessible to global measures alone. These findings reinforce the characterization of gravcat states as genuine quantum resources. Future work could examine how correlated dephasing channels affect the $C_C$/$C_L$ redistribution, and extend the framework to multipartite gravcat systems to probe richer gravitationally induced coherence structures.

\bibliographystyle{unsrtnat}

\end{document}